\documentclass[12pt,english]{article}
\usepackage[english]{babel}
\usepackage{amsmath,amssymb,amscd}
\usepackage{amsfonts}
\usepackage{graphicx}

\makeatletter
\renewcommand\section{\@startsection {section}{1}{\z@}%
                                   {-5.5ex \@plus -1ex \@minus -.2ex}
                                   {2.3ex \@plus.2ex}%
                                   {\normalfont\large\bfseries}}
\renewcommand\subsection{\@startsection{subsection}{2}{\z@}%
                                     {-3.25ex\@plus -1ex \@minus -.2ex}%
                                     {1.5ex \@plus .2ex}%
                                     {\normalfont\bfseries}}

\numberwithin{equation}{section}

\makeatother

\date{September 2012}

\newcommand{\ba}{\begin{align}}

\newcommand{\bea}{\begin{eqnarray}}
\newcommand{\eea}{\end{eqnarray}}
\newcommand{\be}{\begin{equation}}
\newcommand{\ee}{\end{equation}}

\newcommand{\Z}{{\mathbb Z}}
\newcommand{\R}{{\mathbb R}}

\def\Tr{{\rm Tr}}

\newcommand{\cR}{{\cal R }}

\newcommand{\ie}{{\it i.e.~}}
\newcommand{\eg}{{\it e.g.~}}

\newcommand{\bs}{{\bar{s}}}
\newcommand{\bw}{{\bar{w}}}
\newcommand{\bh}{{\bar{h}}}

\newcommand{\elli}{J(q,y)}
\newcommand{\ma}{{\textrm{ch}}}
\newcommand{\ml}{\chi}
\newcommand{\gma}{g}
\newcommand{\gml}{\gamma}

\title{\vspace{-1cm}\begin{flushright}{\small CALT-68-2885, IPMU12-0153, RUNHETC-2012-19}\end{flushright}\vspace{2cm}
\LARGE Modular Constraints on Calabi-Yau Compactifications
}

\author
{
Christoph A.~Keller$^1$\footnote{Address after September 1: NHETC and Department of Physics and Astronomy, Rutgers University,
Piscataway, NJ 08855-0849, USA }\ \footnote{keller@physics.rutgers.edu}\ , Hirosi Ooguri$^{1,2}$\footnote{ooguri@theory.caltech.edu}
\\
\\
$^1$ California Institute of Technology, Pasadena, CA 91125, USA\\
$^2$ Kavli IPMU (WPI), 
University of Tokyo,
 Kashiwa 277-8583, Japan
}

\begin{document}

\maketitle

\begin{center}
{\bf Abstract}
\end{center}

We derive global constraints on the non-BPS sector of 
supersymmetric 2d sigma-models whose target space
is a Calabi-Yau manifold.
When the total Hodge number of the Calabi-Yau threefold is sufficiently large, 
we show that there must be non-BPS primary states whose total conformal weights 
are less than $0.656$. 
Moreover, the number of such primary states grows at least
linearly in the total Hodge number. 
We discuss implications of these results for Calabi-Yau geometry.

\newpage

\section{Introduction}
In this paper we derive constraints on the non-BPS sector 
of two-dimensional supersymmetric sigma-models whose target 
space is a Calabi-Yau manifold. Although we will focus on Calabi-Yau 
manifolds in three complex-dimensions, most of our results are readily 
generalizable to higher dimensions. 

The purpose of this work is two-fold. Over the past few decades, 
remarkable progress has been made in understanding BPS sectors of 
supersymmetric quantum field theories in various dimensions. 
On the other hand, very little is known about properties of their non-BPS sectors. 
Generally speaking, non-BPS sectors of supersymmetric theories 
could be as inaccessible as theories without supersymmetry. 
In this paper, we show that there are ways to take advantage 
of known facts on BPS sectors of two-dimensional 
supersymmetric sigma-models 
to derive non-trivial statements on its non-BPS sector. 

Our second motivation comes from the question of whether there are 
infinitely many topological types of Calabi-Yau manifolds.
In two complex dimensions, there is only one topological type,
which is the K3 surface. 
For three-folds, 
it has been conjectured in \cite{yau} that the number of topological types 
is finite (see also \cite{Reid}). This question
has an obvious relevance for the study of the landscape of string vacua
for example. If the number of topological types is finite, it would 
follow in particular that the Hodge numbers should be bounded. 
This leads to the natural question of whether the
sigma-model shows any distinct behavior when we postulate
the existence of a target manifold with extremely large Hodge numbers. 

Although we are not going to present a definite answer to this question, 
we find that the non-BPS sector of the sigma-model shows
interesting features when the total Hodge number $h_{{\rm total}}= h^{1,1}+ h^{2,1}$
is large. We will prove that when $h_{{\rm total}}$
is sufficiently large, there must be non-BPS primary states whose
total conformal weight $\Delta_{{\rm total}}= \Delta + \bar{\Delta}$ 
is less than $0.655\cdots$. In contrast, $\Delta_{{\rm total}}$
of the first non-trivial BPS states is $1$, so that in particular
it follows that the lowest primary field is non-BPS. Moreover 
we show that the number of such primary states grows at least
linearly in the total Hodge number. 

Note that the value $0.655\dots$ is a universal upper bound for
the lowest non-zero conformal weight, independently of moduli 
of the target Calabi-Yau manifold. The bound can be lower in some 
parts of the moduli space. For example, when the volume of the 
target space becomes large, the lowest eigenvalue of the Laplacian 
becomes small and hits zero in the infinite volume limit.
 In contrast, our bound holds everywhere on the moduli space,
including regions where stringy effects are significant. 

To illustrate this point, let us remind ourselves of a much simpler case, 
namely the bosonic sigma model with target space $S^1$.
In this case, the only moduli of the problem is $R$, the radius of the circle. 
When $R$ is large, Kaluza-Klein states become light, and the lowest 
such state has total weight $\frac{1}{2R^2}$. As we make the radius 
smaller, the Kaluza-Klein states become heavier. On the other hand, 
winding states become lighter, the lightest one having
total weight $\frac{1}{2}R^2$. At the self-dual radius $R=1$,
those two states both have weight $\frac{1}{2}$. 
In this particular example the best possible bound we can find
is thus $\Delta_0=\frac{1}{2}$.

Another interesting fact is that the number of such non-BPS
states with $\Delta_{{\rm total}} < 0.655\dots$ grows linearly in 
the total Hodge number $h_{{\rm total}}$. 
In general we do not know how conformal weights 
are distributed between $0$ and $0.655\dots$, but if they are sufficiently evenly 
distributed, the spectrum will become increasingly dense for
large $h_{{\rm total}}$ and become continuous in the limit of
$h_{{\rm total}}\rightarrow \infty$. One interpretation of this result is
that the target Calabi-Yau manifold is forced to decompactify in this limit. 

Finally our results can also be applied to type~II string compactifications,
where they give constraints on massive single string states. 
Note that all our statements so far refer to the 
partition function before the GSO projection, which will
of course eliminate all tachyonic states. 

\begin{figure}[htbp]
\begin{center}
\includegraphics{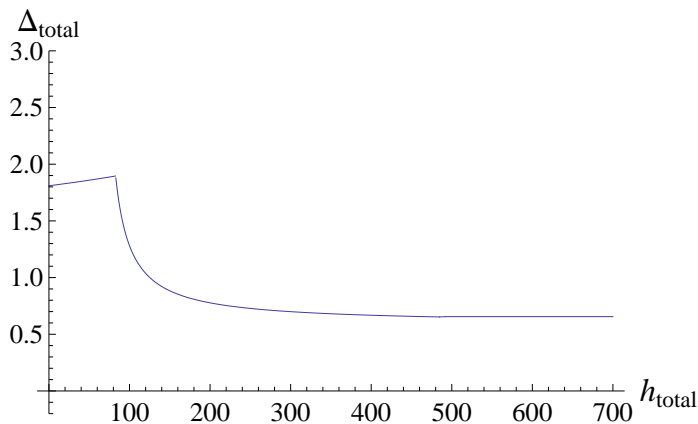}
\caption{Upper bound for the total weight $\Delta_{{\rm total}}=\Delta + \bar{\Delta}$ of the 
lowest lying massive field for a Calabi-Yau threefold as a function of the
total Hodge number $h_{{\rm total}}=h^{1,1}+h^{2,1}$.}
\label{Fig:c9totalconstraints}
\end{center}
\end{figure}

The value $0.655\dots$ that we have quoted so far is not the optimal upper bound. 
In fact, for each 
$h_{{\rm total}}$ greater than 492, we can find a stronger bound.
This is shown in figure~\ref{Fig:c9totalconstraints} above, which for large $h_{{\rm total}}$
asymptotes to $\frac{1}{2}$. Conversely, we can also give bounds for $h_{{\rm total}}$
less than or equal to 492, although those bounds are weaker than the one quoted so far. 
A detailed description of the upper bound of
$\Delta_{{\rm total}}$ as a function of the total Hodge number 
will be given in the main text. 

This paper is organized as follows. 
In section 2, we review the $N=2$ superconformal algebra extended by 
the spectral flow operator. Basic properties of irreducible representations of
this algebra and their characters are described. The Hilbert space of the theory 
can be decomposed into the $\frac{1}{2}$ BPS sector, the $\frac{1}{4}$ BPS sector, 
and the non-BPS sector, corresponding to different representations 
of the extended $N=2$ algebra.
The $\frac{1}{2}$ BPS sector is completely determined by the Hodge number of 
the target Calabi-Yau manifold and the $\frac{1}{4}$ BPS sector is almost determined by its
elliptic genus. 
In section 3, we use modular invariance of the partition function to impose constraints on 
the non-BPS sector. We first illustrate our idea in a generic conformal field theory which may or may
not have supersymmetry. We then combine it with the constraints on the BPS sectors derived in
section 2 to prove the existence of non-BPS states with low conformal weights and to estimate their number.
In the case of Calabi-Yau threefolds, this gives constraints on the 
spectrum if the total Hodge number is $h_{\rm total}\geq 492$. 
In section 4, to obtain constraints also for smaller Hodge numbers, we generalize the medium temperature expansion
introduced in \cite{Hellerman:2009bu,Hellerman:2010qd} and explore consequences of the modular invariance using
differential equations on the partition function. 
We also discuss cases where the worldsheet chiral symmetry is greater than 
the extended $N=2$ superconformal symmetry and when the target space is K3. 
Basic properties of characters of the extended $N=2$ superconformal algebra are 
summarized in Appendix A, and basic properties of elliptic genera are summarized 
in Appendix B.

\section{The extended $N=2$ superconformal algebra}
\subsection{Representations of the extended $N=2$ superconformal algebra}
We will study a CFT realized as a two-dimensional supersymmetric 
non-linear sigma-model whose target space is a Calabi-Yau manifold. 
Note that we will not assume that we are in the large
volume region of the moduli space, \ie that the
geometric description is valid. We only assume that the
moduli space includes a point where
the semi-classical geometric description is valid. 
This assumption provides information on symmetry and
BPS states of the CFT. 

Let us first discuss the symmetry algebra of such a
compactification.
Any kind of Calabi-Yau compactification automatically
has $N=2$ superconformal symmetry for the left and right movers 
on the worldsheet.
In addition it is also invariant under the spectral flow.
The spectral flow by one unit will map
the R and NS sectors onto themselves, and the corresponding
operators will be part of the symmetry algebra.
Geometrically this arises from the fact that a Calabi-Yau $d$-fold has
a holomorphic $(d,0)$ form $\Omega$. This differential form leads
to an operator of dimension $(\frac{d}{2},0)$ in the
spectrum of the theory, which is exactly the spectral
flow operator for one unit of the flow. In the NS-R formalism discussed in
this paper, the spacetime supersymmetry operator for the Calabi-Yau compactification is constructed by
using a square-root of the spectral flow operator, in an appropriate sense. 

The best known example of this enhancement is for K3,
where the spectral flow operator has dimension
$(1,0)$ and enhances
the worldsheet symmetry to the $N=4$ superconformal symmetry,
which includes the $SU(2)$ affine Lie algebra as an $R$ symmetry. 
The representation theory of the $N=4$ superconformal algebra was
studied in \cite{Eguchi:1987sm, Eguchi:1987wf} and applied to
K3 compactifications in \cite{Eguchi:1988vra}. 
For higher dimensional Calabi-Yau manifolds, 
the combination of the $N=2$ superconformal symmetry and the
spectral flow symmetry also generates a larger symmetry algebra, which we will
call the extended $N=2$ superconformal algebra. Its representations have been
investigated in \cite{Odake:1988bh,Odake:1989dm}. 
Here we will only describe results directly relevant to this section.
For detailed expressions of the characters see
appendix~\ref{app:N2char}.

Because of the spectral flow symmetry, there is a natural relation between the NS
and R sectors of the theory, and 
 we can focus on the NS sector without loss of generality. 
There are two types of representations of the extended $N=2$ superconformal algebra: 
chiral (also known as {\it massless}) 
and non-chiral (also known as {\it massive}). 
A non-chiral representation is a highest weight representation whose highest 
weight state $| \Delta, Q \rangle$
is annihilated by all positive modes of the $N=2$ generators
\be
L_n| \Delta, Q \rangle = G_r^\pm | \Delta, Q \rangle = J_n | \Delta, Q \rangle = 0,\quad
n= 1, 2, \dots, \ r= 1/2,3/2,  \dots, 
\ee
and is an eigenstate of $L_0$ and $J_0$ with
\be
L_0 | \Delta, Q \rangle = \Delta | \Delta, Q \rangle , ~ J_0 | \Delta, Q \rangle = Q | \Delta, Q \rangle,
\ee
where $L_n$, $G_r^\pm$, and $J_n$ are generators of the $N=2$ superconformal algebra in the standard notation. 
Similar conditions hold for the modes of the spectral flow
operators.
 
A chiral representation is also a highest weight representation, but
it is {\it shorter} since it obeys the additional
 condition,
\be 
   G_{-1/2}^+ | \Delta, Q \rangle = 0~~{\rm or}~~ G_{-1/2}^- | \Delta, Q \rangle = 0.
\ee
In this case it follows from the $N=2$ commutation relations that
the conformal dimension $\Delta$ and the U(1) charge $Q$ are related as
\be
   \Delta = \frac{1}{2} |Q|\ . 
   \ee

The character of a representation $\cR$ is defined as, 
\be
\chi_\cR = \Tr_\cR y^{J_0}  q^{L_0-\frac{c}{24}}\ .
\ee
For a Calabi-Yau $d$-fold the central charge is $c=3d$ 
and the extended $N=2$ superconformal
algebra has $(d-1)$ massive representations
\be
\ma^Q_\Delta\qquad Q=0, \ldots, d-2 \ .
\ee
For future convenience we will also define $\ma^Q=q^{-\Delta}\ma^Q_\Delta$,
which does not depend on $\Delta$ as can be seen from
the expressions in appendix~\ref{app:N2char}.
There are $d$ chiral representations,
\be
\ml^Q\qquad Q=0,\ldots, d-1 \\ ,
\ee
where $\ml^0$ is the vacuum representation.
Because of the spectral flow symmetry, it is convenient
to extend the range of $Q$ in parameterizing characters as,
$\ml^{Q+d}=\ml^{Q}$ so that $Q$ can take any integer value.
Explicit expressions for the characters can be found
in appendix~\ref{app:N2char}.
The Witten index of the representations is given by
\be\label{WittenIndex}
ind = \left\{\begin{array}{cl}0&\textrm{massive}\\
(-1)^{d-Q}& \textrm{massless}\ Q=1,\ldots, d-1\\
1+(-1)^d&\textrm{massless}\ Q = 0 \end{array}\right.
\ee
When the massive representations reach the unitarity
bound $\Delta=\frac{1}{2}|Q|$, they become reducible and decompose
into massive representations as
\begin{eqnarray}
&&\ma^Q_{Q/2}= \ml^Q+\ml^{Q+1},~~ Q>0\nonumber\\
&&\ma^{-Q}_{Q/2}= \ml^{-Q}+\ml^{-Q-1},~~ Q>0\label{massivetomassless}\\
&&\ma^0_0= \ml^0 +\ml^1+\ml^{-1},~~ Q=0\nonumber
\end{eqnarray}
Note that in these combinations the Witten indices cancels out. 

\subsection{Calabi-Yau partition functions}\label{ss:CYpf}

The Hilbert space of the sigma-model is a sum of representations of the
extended superconformal algebras for the left and right-movers. 
Thus, the partition function is a sum of products of the
$N=2$ characters described in the above. 
Because of the spectral flow symmetry, we will focus on the NS-NS sector. 
It will be useful to decompose the partition function into
different BPS sectors as
\be
Z=Z_{\ml\bar{\ml}} +Z_{\ma\bar{\ml}}+Z_{\ml\bar{\ma}} + Z_{\ma\bar{\ma}}\ .
\ee
Let us now discuss the terms separately.

\bigskip
\noindent (1) \underline{Hodge numbers and the $\frac{1}{2}$ BPS sector}

\medskip 
\noindent
Here $Z_{\ml\bar{\ml}}$ only contains chiral representations for both left and right movers,
representing the $\frac{1}{2}$ BPS sector,
\be
Z_{\ml\bar{\ml}} = \ml^0\bar{\ml}^0+ \sum_{Q, \bar{Q} \neq 0} N_{Q\bar{Q}} \ml^Q\bar{\ml}^{\bar{Q}}\ .
\ee
The degeneracy of this sector is determined by the topology of the Calabi-Yau, 
as the coefficients $N_{Q\bar{Q}}$ are fixed by the Hodge numbers of the target space. 
For example, for a Calabi-Yau 3-fold, 
\be
Z_{\ml\bar{\ml}} = \ml^0\bar{\ml}^0+ h^{1,1}( \ml^1\bar{\ml}^{1} + \ml^{-1}\bar{\ml}^{-1})
+ h^{2,1} ( \ml^1\bar{\ml}^{-1} + \ml^{-1}\bar{\ml}^{1}).
\ee
This is also the reason why we did not include terms of the form $\ml^{Q=0} \bar{\ml}^{\bar{Q}}$
or  $\ml^{Q} \bar{\ml}^{\bar{Q}=0}$, since they contribute to the partition function only
when the target space is a torus.  

\bigskip
\noindent (2) \underline{$\frac{1}{4}$ BPS sector and the elliptic genus}

\medskip
\noindent
The mixed terms or $\frac{1}{4}$ BPS states are contained in $Z_{\ma\bar{\ml}}+Z_{\ml\bar{\ma}}$,
\be
Z_{\ma\bar{\ml}}= \sum_{Q,\bar Q, \Delta} N^\Delta_{Q,\bar{Q}} \ma_\Delta^Q \bar\ml^{\bar Q}.
\ee
This sector is essentially determined by the elliptic genus defined by
\be
J(q,y)= \bar q^{\frac{d}{8}}Z(q,\bar q; y,\bar{y} = -\bar{q}^{1/2}) ,
\ee
where we used the spectral flow to shift the right-mover to the R sector,
while keeping the left-mover in the NS sector.\footnote{For notational
convenience we have kept the left movers in the NS sector, whereas usually they
are taken to be in the R sector. It is straightforward to 
spectrally flow from our definition to the standard definition, in particular
when identifying with weak Jacobi forms.}
By supersymmetry it does not depend on $\bar q$.
In the R sector, each character is replaced by the Witten index (\ref{WittenIndex}).
Thus the elliptic genus can be expanded in terms of the characters of the extended $N=2$ algebra as,
\begin{eqnarray}
J(q,y)&=& \left( 1 + (-1)^{d}\right) \left( \ml^0 + \sum_Q N^\Delta_{Q,0} \ma^Q_\Delta\right) + \\
&& + \sum_{\bar{Q} \neq 0}(-1)^{\bar Q} \left[ \sum_{Q\neq 0} N_{Q\bar{Q}}\left((-1)^d \ml^Q + \ml^{-Q} \right) 
+ \sum_{Q. \Delta} N^\Delta_{Q,\bar Q} \left((-1)^d \ma^Q_\Delta + \ma^{-Q}_\Delta\right) \right] . \nonumber
\end{eqnarray}
Note that here and in the following by a slight abuse of notation
we use the same symbols $\ml$ and $\ma$ for the characters 
of the different 
s.

If we take the elliptic genus to be in the RR spin structure, 
then it transforms nicely under modular transformations.
In addition, it has periodicity under $y \rightarrow qy$ because of the spectral flow symmetry. 
More precisely it is a weak Jacobi form of weight 0 and index $d/2$.
We describe weak Jacobi forms in more detail in appendix~\ref{app:ellGenus}.
What is important here is that the space $J_{0,d/2}$ 
of such forms is finite dimensional.
For small $d$ the elliptic genus $\elli$ is fixed by $Z_{\ml\bar\ml}$.\footnote{For larger $d$
one also needs to fix some of the lower terms of $Z_{\ma\bar\ml}$. See \cite{Gaberdiel:2008xb} for
a detailed discussion of which terms determine a weak Jacobi form uniquely.} From $J(q,y)$ we can
thus read off the coefficients of $\ma^Q_\Delta$, 
which fixes the combinations $\sum_{\bar{Q}} (-1)^{\bar Q}N^\Delta_{Q\bar{Q}}$.

Let us now discuss the cases $d=2$ and $d=3$ explicitly.

\bigskip
\noindent K3: 

\medskip
The elliptic genus of K3 can be expanded in terms of characters as
\be
J(q,y) = 2 \ml^0 - 20 \ml^1 - 2 \sum_{n=1}^\infty N_n \ma^{Q=0}_{\Delta=n}\ .
\ee
Note that in this case the corresponding weak Jacobi form is unique up to overall normalization.
The fact that there is a unique vacuum state
forces the coefficient of $\ml^0$ to be $2$, which fixes the normalization
and thus the full elliptic genus. 
In particular one can derive from this that the coefficient of $\ml^1$ is $20$, 
which reproduces that $h^{1,1}=20$ for K3 \cite{Eguchi:1988vra}. 

In \cite{Ooguri}, it was conjectured that all the coefficients $N_n$ are non-negative integers,
based on the following reasoning. These coefficients are related to the multiplicities 
$N^\Delta_{Q,\bar{Q}}$ as
\be
  N_n = N^{\Delta = n}_{Q=0, \bar{Q}=1} - N^{\Delta=n}_{Q=0, \bar{Q}=0}\ . 
\ee
If for some $n$ it were the case that $N_n < 0$ it would follow that $N^{\Delta=n}_{Q=0, \bar{Q}=0} > 0$. 
This would imply that there is a $\frac{1}{4}$ BPS state of dimension $(n, 0)$, which 
would enhance the $N=4$ superconformal algebra of the sigma-model to a larger chiral algebra everywhere 
in the moduli space of K3. Such an enhancement is contrary to the expectation that the generic chiral algebra
of the K3 sigma-model is simply the $N=4$ superconformal algebra. An explicit computation
up to $n=100$ gave indeed positive integer values for the $N_n$ \cite{Ooguri}.
The first few values are given by 
\be
N_{1}=90\ , \qquad
N_{2}=462\ , \qquad
N_{3}=1540\ , \qquad
N_{4}=4554\ ,  \ldots\ .
\ee
Moreover, the behavior of $N_n$ for $n\gg 1$ was estimated using the modular properties of the elliptic genus 
and of the $N=2$ characters, and the result turned out to be consistent with the expectation that
$N_n > 0$ for all $n$. 

More recently, it was pointed out in \cite{Eguchi:2010ej} that the coefficients $N_n$ are related in a simple manner to
dimensions of representations of the largest Mathieu group $M_{24}$, suggesting that the Mathieu group acts non-trivially on
the elliptic cohomology of K3. Further evidence in favor of this conjecture has been found, for example, in \cite{Gaberdiel}.

To summarize, if one assumes that there is no enhancement of the chiral algebra, namely $N^{\Delta=n}_{Q=0, \bar{Q}=0}=0$,
then the multiplicities of the $\frac{1}{4}$ BPS sector of K3 are completely determined by the elliptic genus.

\bigskip
\noindent Calabi-Yau threefold:

\medskip

The elliptic genus in this case is given by,
\be
J=(h^{1,1}-h^{2,1})(\ml^1-\ml^{-1})+
\sum_{Q=0,1 } \sum_{n=1}^\infty (N^{\Delta=n}_{Q, 1} - N^{\Delta=n}_{Q, -1}) \ma^Q_n .
\label{threefoldgenus}
\ee
The space of relevant Jacobi forms is again one-dimensional and
the elliptic genus should be proportional to the unique weak Jacobi form $\phi_{0,3/2}$
of weight $0$ and index $3/2$ defined in appendix~\ref{app:ellGenus}.
Matching with the Euler characteristic of the Calabi-Yau manifold gives \cite{Gritsenko:1999fk} 
\be
J= (h^{1,1}-h^{2,1})\phi_{0,3/2}\ .
\ee
Rather surprisingly it turns out that 
\be
\ml^1- \ml^{-1}=\phi_{0,3/2}\ .
\ee
Thus, the elliptic genus is entirely given by the first term in the right-hand side of (\ref{threefoldgenus}),
and the second term should be zero, namely, 
\be
   N^\Delta_{Q, \bar{Q}=1} = N^\Delta_{Q, \bar{Q}=-1}. 
\ee
Taking this into account, the $1/4$ BPS part of the partition function is given by,
\begin{eqnarray}
Z_{\ma\bar{\ml}} &=&  \sum_{Q=0,1} \sum_{n=1}^\infty \left( N^{\Delta=n}_{Q, \bar{Q}=0} \ma_n^Q 
\bar{\ml}^{0}  +    N^{\Delta = n}_{Q, 1} \ma_n^Q (\ml^1 + \ml^{-1})\right) \nonumber \\
&=&    \sum_{Q=0,1} \sum_{n=1}^\infty  \left( N^{\Delta=n}_{Q, \bar{Q}=0} \ma_n^Q 
\bar{\ml}^{0}  +   N^{\Delta = n}_{Q, 1} \ma_n^Q\ma^1_{1/2}\right),
\end{eqnarray}
where we used
\be
  \ml^1 + \ml^{-1} = \ma^1_{1/2},
\ee
which holds for $d=3$. 
The second term in the right hand side can then be regarded as a part of the 
non-BPS sector of the partition function $Z_{\ma\bar{\ma}}$.
Furthermore, if we assume that there is no enhancement of the chiral 
symmetry beyond the extended $N=2$ superconformal algebra,
we should set $N^n_{Q, \bar{Q}=0}=0$. Under this assumption, the 
$\frac{1}{4}$ BPS part of the partition function is fixed and in fact is zero, 
\be
  Z_{\ma\bar{\ml}} =0. 
\ee

\bigskip
We have seen that, for two and three complex-dimensions, if we assume no enhancement of the chiral symmetry, 
the elliptic genus fixes the $\frac{1}{4}$ BPS sector. The situation in higher dimensions is different and will be discussed later. 
In the next section, we will analyze the structure of the non-BPS part of the partition function under this assumption. 
Later, we will relax this condition, but will find that the results are not qualitatively changed.

\section{Modular invariance: the direct way}

\subsection{A first attempt}

To begin with, let us consider a partition function of a generic unitary conformal field theory.  
Define $\tau=ix$ with $x\in\R$. 
The modular invariance under the $S$ transform implies that
\be \label{maindiff}
Z(x)-Z(1/x)=0\ .
\ee
Suppose $x\geq 1$.
The central observation is that a state
of total weight $\Delta_{{\rm total}}>\frac{c}{12}$ gives a negative contribution
to the left-hand side of (\ref{maindiff}) since
\be \label{gcontribution}
e^{-2\pi (\Delta_{{\rm total}}-c/12)x}-e^{-2\pi(\Delta_{{\rm total}}-c/12)/x}<0\ .
\ee
Assuming the theory is unitary, only states with $\Delta_{{\rm total}}<\frac{c}{12}$ give positive
contributions, 
and thus at least one such state must exist in order for (\ref{maindiff}) to hold. 

However, this condition by itself is trivial since any unitary conformal field theory
contains the vacuum state, which certainly satisfies the inequality $\Delta_{{\rm total}}=0<\frac{c}{12}$.
To achieve a non-trivial statement, we need to have more information
on the spectrum of the theory. More precisely we want to consider
theories that have several states that give positive contributions
to (\ref{maindiff}), which need to be compensated by additional
low lying states other than the vacuum.

In view of applying this observation to theories with fermions,
note that for (\ref{maindiff}) we have not assumed that the partition function is 
invariant under the full modular group, but only under $S$. 
Thus, we can stay within the NS-NS
spin structure. In the following, we
will always work with the NS sector characters.

\subsection{Using Hodge numbers}

Our strategy is to combine the above observation with our knowledge on the 
$\frac{1}{2}$ BPS sector and the $\frac{1}{4}$ BPS sector to derive non-trivial constraints on the non-BPS sector.
We will focus on the case when the target space Calabi-Yau has three complex dimensions 
and discuss higher dimensional cases later.  

The fundamental observation is as follows. 
While the vacuum representation $\ml^0 \bar\ml^0$ gives a positive contribution
to the left-hand side of (\ref{maindiff}),
the other $\frac{1}{2}$ BPS representations, namely $h_{1,1} (\ml^1\bar\ml^1+ \ml^{-1}\bar\ml^{-1})$ and 
$h_{2,1} (\ml^1\bar\ml^{-1}+ \ml^{-1}\bar\ml^{1})$, give negative contributions.
This can be proven using the characters of these representations. 
Therefore, if $h_{1,1}$ and $h_{2,1}$ are sufficiently large, the $\frac{1}{2}$ BPS sector as a whole 
gives a negative contribution to the left-hand side of (\ref{maindiff}), 
\be
 Z_{\ml\bar\ml}(\tau=ix, \theta=0) - Z_{\ml\bar\ml}(\tau=i/x, \theta=0) < 0. 
\ee
As we saw in the previous section, the $\frac{1}{4}$ BPS sector is absent if we assume that
there is no enhancement of the chiral algebra. Therefore, 
in order to satisfy the equality (\ref{maindiff}) for large Hodge numbers, the non-BPS sector should contribute 
positively, 
\be
 Z_{\ma\bar\ma}(\tau=ix, \theta=0) - Z_{\ma\bar\ma}(\tau=i/x, \theta=0) >0. 
\ee
However, we will find that only those representations $\ma_\Delta^Q \bar\ma_{\bar\Delta}^{\bar Q}$
with small enough $\Delta + \bar{\Delta}$ can give positive contributions. 
This then leads to an upper bound on the lowest value of  $\Delta + \bar{\Delta}$ .

To evaluate the contribution of each representation to the left-hand side of  (\ref{maindiff}), 
let us define
\begin{eqnarray}
\gml^{Q\bar Q}(x)&=& \ml^{Q}(x)\ml^{\bar Q}(x)-\ml^{Q}(x^{-1})\ml^{\bar Q}(x^{-1})\\
\gma^{Q\bar Q}_{\Delta\bar\Delta}(x)&=& \ma^{Q}_\Delta(x)\ma^{\bar Q}_{\bar{\Delta}}(x)-\ma^{Q}_\Delta(x^{-1})\ma^{\bar Q}_{\bar{\Delta}}(x^{-1})
\end{eqnarray}
In total (\ref{maindiff}) yields
\be
\gml^{00}(x)+2(h^{1,1}+h^{2,1})\gml^{11}(x)+\sum N_{Q\bar Q}^{\Delta\bar{\Delta}}\gma^{Q\bar Q}_{\Delta\bar{\Delta}}(x)= 0\ .
\ee
Note that from (\ref{gcontribution}) it follows
that we can easily bound $g$ and $\gamma$ from above
by evaluating a finite number of terms, since all but
finitely many terms are negative. 
In particular it follows immediately that $\gml^{11}(x)<0$.

To find a bound, we need to establish for which values
of $\Delta_{\rm total}$ massive representations give positive contributions.
Let us define 
\be \label{sup}
S(x)=\sup(\Delta+\bar{\Delta}:\gma^{Q\bar Q}_{\Delta\bar{\Delta}}(x)>0, Q, \bar Q=0,1)\ .
\ee
Note that from (\ref{gcontribution}) it follows immediately that $S(x)\leq \frac{3}{4}$.

We can use this to give a lower bound for the number of states
with total weight less than $S(x)$ by writing
\be
\sum_{\Delta+\bar{\Delta}\leq S(x)} N_{Q\bar Q}^{\Delta\bar{\Delta}}\gma^{Q\bar Q}_{\Delta\bar{\Delta}}(x)\geq
-\gml^{00}(x)-2(h^{1,1}+h^{2,1})\gml^{11}(x)\ .
\ee
For fields with $\Delta \leq \frac{3}{4}$ we can
bound
\be\label{boundcharacter}
\gma^{ab}_{h\bh}(x)\leq e^{2\pi\frac{3}{4}x}+4e^{2\pi\frac{1}{4}x}
-e^{2\pi\frac{3}{4}x^{-1}}-4e^{2\pi\frac{1}{4}x^{-1}}\ .
\ee
To see this, we note that only the first couple of terms in the expansion of each character
give positive contributions to $g_{\Delta\bar{\Delta}}^{Q\bar Q}$. For
example, 
\be\label{derivebound}
\gma^{00}_{\Delta \bar{\Delta}}(x)\leq \left(e^{2\pi(\frac{3}{4}-\Delta_{{\rm total}})x}-e^{2\pi(\frac{3}{4}-\Delta_{{\rm total}})x^{-1}}\right)
+\left(4e^{2\pi(\frac{1}{4}-\Delta_{{\rm total}})x}-4e^{2\pi(\frac{1}{4}-\Delta_{{\rm total}})x^{-1}}\right)\ .
\ee
We then arrive at (\ref{boundcharacter}) by noting that the right-hand side is a decreasing function in $\Delta_{{\rm total}}$
and thus one gets the universal bound by setting $\Delta_{{\rm total}}=0$ on the right-hand side. 
A similar argument and remembering that $\Delta\geq\frac{1}{2}$ for $\gma^{10}_{\Delta\bar\Delta}$ shows that 
(\ref{boundcharacter}) is also satisfied in this case.

In total we have thus found the following bound on the number of low-lying states:
\be\label{Nbound}
\# \{ \textrm{non-BPS~primaries~with~}\Delta_{{\rm total}} \leq S(x)\} 
\geq \frac{2(h^{1,1}+h^{2,1})|\gml^{11}(x)| -\gml^{00}(x)}{e^{2\pi\frac{3}{4}x}+4e^{2\pi\frac{1}{4}x}
-e^{2\pi\frac{3}{4}x^{-1}}-4e^{2\pi\frac{1}{4}x^{-1}}}.
\ee
We used the absolute value symbol as $|\gamma^{11}|= -\gamma^{11}$ since 
$\gamma^{11}$ is always negative. 

Note that we have a bound for each $x \geq 1$, so in principle
we can try to choose $x$ optimally to find the best bound. 
Let us explore this in detail.

\subsection{Upper bounds}
We will leave finding the optimal value of $x$ for future work. Instead we will
only consider two special values for $x$ and discuss the resulting bounds (\ref{Nbound}). 
The first value we want to consider is $x \rightarrow 1$. 
In this case the series expansion of the characters 
converges very quickly, so that it is straightforward to obtain
a numerical value for $S(1)$, namely
\be
 S(x=1) = 0.655 \dots\ . 
\ee
Moreover we have 
\begin{eqnarray}
&&\frac{\gml^{00}(x)}{2|\gml^{11}(x)|}\rightarrow 492.6\dots,\nonumber \\
&& \frac{ e^{2\pi\frac{3}{4}x}+4e^{2\pi\frac{1}{4}x}
-e^{2\pi\frac{3}{4}x^{-1}}-4e^{2\pi\frac{1}{4}x^{-1}}}{2|\gml^{11}(x)|}
\rightarrow 522.0\dots,~~~~(x \rightarrow 1),
\end{eqnarray}
so that we find
\be
\# \{ \textrm{non-BPS~primaries~with~}\Delta_{{\rm total}} \leq 0.655 \dots \}
\geq \frac{1}{522.0 \dots}\left( h^{1,1} + h^{2,1} - 492.6 \dots\right). 
\ee
This gives non-trivial constraints when the total Hodge number is greater than 492. 

It is interesting to note that there are Calabi-Yau manifolds with such Hodge numbers, 
but they tend to be on the upper end of
known constructions. For example, the largest value of $h_{2,1}$ for any elliptically fibered Calabi-Yau threefold 
is 491 \cite{taylor}.  The elliptic threefold at the upper bound of $h_{2,1}$ has $h_{1,1}=11$, thus the total 
Hodge number exceeds $492$.  Scanning hypersurfaces
in toric manifolds suggests that the maximum value of $(h_{1,1}+h_{2,1})$ for this class of Calabi-Yau
manifolds is also around $500$. 

Let us consider the other extreme of $x \rightarrow \infty$. In this limit, 
\be\label{Sxinfinity}
 S(x \rightarrow \infty) = \frac{1}{2},
\ee
so the bound on $\Delta_{{\rm total}}$ is better. 
To obtain (\ref{Sxinfinity}), note that that the asymptotic
growth of the characters is given by the Cardy formula, so that
\be
\gma^{Q\bar{Q}}_{\Delta\bar{\Delta}} \sim_{x\rightarrow\infty} e^{-2\pi(\Delta_{\rm total}-\frac{3}{4})x} - e^{2\pi \frac{c_{eff}}{12}x}\ ,
\ee
where $c_{eff}=3$ is the effective central charge of 
the extended $N=2$ algebra. 
However, 
\be\label{Nbound1}
\# \{ \textrm{non-BPS~primaries~with~}\Delta_{{\rm total}} \leq 1/2  \}
\geq 2 e^{-2\pi\frac{1}{2}x}\left(h^{1,1} + h^{2,1}\right) -1,~~~~ x \rightarrow \infty,  
\ee
in this limit. Thus, we need increasingly large ($\sim e^{2\pi\frac{1}{2}x}$) total Hodge number in order to 
use this bound for $x \rightarrow \infty$.

\subsection{Generalization to Calabi-Yau $d$-folds}\label{ss:dfold}

Let us briefly discuss how these methods generalize to 
other dimensions. 
For K3 the Hodge numbers are fixed, and it turns out
that the sum of the 1/2 and 1/4 BPS sectors already gives a positive contribution
for (\ref{maindiff}). 
Thus this method does not give an upper bound for the lowest conformal weight
in $Z_{\ma\bar\ma}$. 

On the other hand, we can generalize
(\ref{Nbound}) to higher dimensions. 
By using the reflection symmetry of the characters, 
\be
\gml^{Q,\bar Q}(x)=\gml^{Q,\bar Q}(x)=\gml^{d-Q,\bar Q}(x)=\gml^{Q,d-\bar Q}(x)\ ,
\ee
we can express the contribution of the massless-massless fields in (\ref{Nbound}) as 
\be
\label{generald}
\gml^{00}(x)+\sum_{Q,\bar Q=1}^{\lfloor d/2\rfloor}
2(h^{Q,\bar Q}+h^{d-Q,\bar Q})\gml^{Q\bar Q}(x)\ ,
\ee
when $d$ is odd. There is a similar formula for $d$ even. 

\begin{figure}[htbp]
\begin{center}
\includegraphics[width=0.4\textwidth]{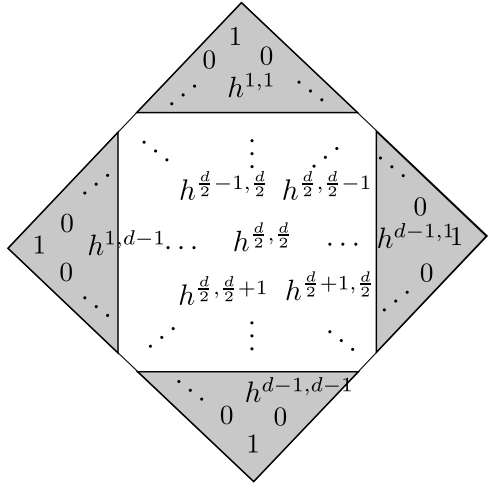}
\caption{The Hodge diamond of a Calabi-Yau of dimension $d$.
The unshaded and shaded regions give positive and negative contributions 
to the right-hand side of
(\ref{higherCY}) respectively.}
\label{Fig:HodgeDiamon}
\end{center}
\end{figure}

The question is then which $\gamma^{Q\bar Q}$ are still negative. 
When $d=3$, the vacuum contribution $\gamma^{00}$ is positive and
the contributions of the chiral fields $\gamma^{11}$ are negative.
The situation is more complicated for higher $d$.
For concreteness consider the limit of $x\rightarrow\infty$. 
In this case we can repeat the argument that leads to
(\ref{Sxinfinity}) and find that
$\gamma^{Q,\bar Q}$ is positive 
for
\be \label{fourcorners}
Q+\bar Q < \frac{d}{2}-1\ .
\ee
When $d=3,4,5$, only the vacuum satisfies this condition, 
and all the other $\gml^{Q\bar Q}(x)$ with $Q, \bar{Q} \neq 0$ give negative
contribution. In these cases, our analysis carries over and 
we get a non-trivial bound on 
$\Delta_{{\rm total}}$ if any one Hodge number is large enough. 
In contrast,  for $d > 5$, terms with small values of
$(Q, \bar Q)$, \ie Hodge numbers near the four corners of the Hodge diamond,
give positive contribution. 
More precisely, the constraint on the low lying fields reads
\begin{multline}\label{higherCY}
\sum_{\Delta+\bar{\Delta}\leq S(x)} N_{Q\bar Q}^{\Delta\bar{\Delta}}\gma^{Q\bar Q}_{\Delta\bar{\Delta}}(x)\geq
-|\gml^{00}(x)|\\-\sum_{Q+\bar Q < \frac{d}{2}-1}^{\lfloor d/2\rfloor}
2(h^{Q,\bar Q}+h^{d-Q,\bar Q})|\gml^{Q\bar Q}(x)|
+\sum_{Q+\bar Q \geq \frac{d}{2}-1}^{\lfloor d/2\rfloor}
2(h^{Q,\bar Q}+h^{d-Q,\bar Q})|\gml^{Q\bar Q}(x)|\ .
\end{multline}
This is only non-trivial when the right hand side is positive.
This can be achieved by increasing the Hodge numbers
close to the center of the Hodge diamond, or more precisely, 
in the unshaded region in 
Fig.~\ref{Fig:HodgeDiamon}. On the other hand
if the Hodge numbers near the corners are big enough, 
the right hand side will be negative.
Note that the shaded region tends to get bigger we
consider values of $x$ other than $x\rightarrow\infty$,
since the condition (\ref{fourcorners}) for
positive contributions becomes weaker.

\section{Differential operators}

\subsection{Differential operators and the medium temperature expansion}
In the previous section we used modular invariance of the partition function
to obtain bounds on the number of low lying massive representations.
This method was straightforward to implement, but as we saw
it only works for relatively large Hodge numbers.
We shall now generalize a method
introduced in \cite{Hellerman:2009bu,Hellerman:2010qd}
which is applicable also for smaller Hodge numbers.

In the NS-NS sector, the partition function,
\be
Z= \Tr e^{2\pi i \tau (L_0-c_L/24)}e^{-2\pi i \bar \tau (\bar L_0-c_R/24)}
e^{2\pi i z J_0} e^{-2\pi i \bar z \bar J_0}\ ,
\ee
transforms under the $S$ modular transformation as,
\be \label{charTrafo}
Z(-1/\tau, z/\tau) \mapsto e^{2\pi i \frac{c_L}{6} \frac{z^2}{\tau}} 
e^{-2\pi i \frac{c_R}{6} \frac{\bar z^2}{\bar \tau}}Z(\tau, z)\ .
\ee
To explore consequences of this equality, we 
use the medium temperature expansion developed in \cite{Hellerman:2009bu,Hellerman:2010qd}.
This means that we expand the partition function 
around the fixed point, $\tau =i, z=0$, of the $S$ transformation.
It is convenient to introduce a new set of coordinates, 
\be
\tau = i e^s\ , \qquad z = e^{\frac{s}{2}}w\ ,
\ee
which have the very simple transformation properties
\be \label{Strafo}
s \mapsto -s\ , w \mapsto -iw\ ,
\ee
and analogously for $\bar \tau$ and $\bar z$.
Because of (\ref{charTrafo}), $Z$ is not quite
invariant under the $S$ transformation.
We therefore define the invariant partition
function
\be \label{hatdef}
\hat Z(s,w)= e^{\pi \frac{c_L}{6} w^2}e^{\pi \frac{c_R}{6} \bar w^2} Z(s,w)\ ,
\ee
where in our case again $c_L=c_R=3d$.
Since $\hat Z$ is invariant under (\ref{Strafo}),
it must satisfy the identity
\be \label{Sinvariance}
\hat Z(s,w) = \hat Z(-s,-iw)\ .
\ee
Note that for $w=0$ we recover the original condition of \cite{Hellerman:2009bu}.

The condition (\ref{Sinvariance}) is the central identity which
we will work with. 
Expanding $\hat Z$ as a Taylor series in $s, \bs,w,\bar w$, 
we can only encounter monomials which are 
invariant under (\ref{Strafo}). Since all other monomials 
must vanish, it follows that the corresponding derivatives at
the origin must vanish. We can thus associate
to each such monomial a differential operator $D$ which must 
annihilate the partition function when evaluated at the origin,
\be
D\hat Z(s,\bs,w,\bar w)|_{\bs=s=w=\bar w=0}=0\ .
\ee
Here $D$ is some monomial in the operators $\partial_s,\partial_w,\partial_\bs,\partial_\bw$.
Note that terms which are odd in the total number of $w$ and $\bar w$
will vanish automatically by charge conjugation, but that the terms will
lead to non-trivial conditions.

The basic strategy is thus to take linear combinations
of such differential operators. These operators
annihilate the partition function, but they 
do not annihilate the characters of the various
representations individually, since those are not modular invariant
by themselves. 

In what follows the action of $D$ on massive representations
will be central.
As pointed out before, the character of a massive representation 
is of the form $\hat Z^{Q\bar Q}_{\Delta\bar\Delta}=q^{\Delta}\bar q^{\bar\Delta}\ma^Q\bar\ma^{\bar Q}$, 
where $\ma^Q$ does not depend on $\Delta$. We can thus define
\be
P_D(\ma^Q_\Delta\bar\ma^{\bar Q}_{\bar{\Delta}})
:=e^{2\pi(\Delta+\bar{\Delta})}D\hat Z^{Q\bar Q}_{\Delta\bar\Delta}\bigr|_0\ ,
\ee
which is a polynomial in $\Delta$ and $\bar{\Delta}$,
whose degree is given by the degree of $D$ in $\partial_{s,\bs}$.

\subsection{Constraints on massless representations}
We will now use our differential operators to see
what constraints modular invariance imposes
on the spectrum. As a warm up, let us try to see
if we can find additional constraints on the massless spectrum.
As we have seen above, the elliptic genus only determines
certain linear combinations of the $N^\Delta_{Q\bar Q}$,
but not all the coefficients themselves.
A priori it thus seems possible to obtain more information.

To do this, let us construct a differential operator $D$
that gives zero for all massive representations
\be
P_D(\ma^Q_\Delta\bar\ma^{\bar Q}_{\bar{\Delta}})=0 \qquad Q,\bar Q=0,\ldots,d-2\ .
\ee
Constructing such an operator is not as difficult
as it may seem. Let us concentrate on holomorphic operators for the moment.
We know that operators which contain a single
derivative in $s$ give linear polynomials,
\be
P_D(\ma^Q_\Delta)=\alpha^Q h +\beta^Q\ .
\ee
A suitable linear combination $D$ of the operators
\be
\partial_s \partial_w^{4i}\ , \qquad \partial_w^{4i+2}\ ,\qquad i=0,\ldots d-1
\ee
will thus annihilate all $d-1$ massive characters. 
It then follows that
\be
0=D(\hat Z)=D(\hat Z_{\ml\bar{\ml}}) + D(\hat Z_{\ma\bar{\ml}})+D(\hat Z_{\ml\bar{\ma}}) \ .
\ee
which indeed seems to impose a constraint on the
chiral representations.
Unfortunately this does not fix the $N^\Delta_{Q\bar{Q}}$, since
we can always use relations such as (\ref{massivetomassless})
to change them by adding massive fields. Since $D$
annihilates massive fields, it is oblivious to such
changes, which means that any condition can only
involve the combinations $\sum_{\bar{Q}} (-1)^{\bar Q}N^\Delta_{Q\bar{Q}}$
which also appear in the elliptic genus. In other words
we will only ever recover conditions that we could
also obtain from the elliptic genus.

\subsection{Constraints on massive representations}
Let us now turn to massive representations, where we are indeed 
able to find interesting constraints.
Here we will turn around the above strategy.
Assume that we know
enough about $Z_{\ma\bar{\ml}}$ that we can construct an
operator $D$ such that\footnote{If the elliptic genus does not allow
$Z_{\ma\bar{\ml}}$ to be fixed,
then we can always try to bound its contribution to get an inequality instead.}
\be
D(\hat Z_{\ml\bar{\ml}}) + D(\hat Z_{\ma\bar{\ml}})+D(\hat Z_{\ml\bar{\ma}}) = 0\ .
\ee
This is again simple to do, since for any two
operators we can get such a $D$ as a suitable linear combination.
It follows thus that also $D(\hat Z_{\ma\bar{\ma}})=0$.
Take the highest derivative in $s,\bs$ to be $(-\partial_{s+\bs})^n$.
As pointed out above, $P_D(\ma^Q_\Delta\bar\ma^{\bar Q}_{\bar \Delta})$
is then a polynomial of degree $n$ in $\Delta_{\rm total}$, with
the leading term
\be
P_D(\ma^Q_\Delta\bar\ma^{\bar Q}_{\bar{\Delta}})=
(2\pi\Delta_{total})^n\ma^Q(e^{-2\pi},1)\bar\ma^{\bar Q}(e^{-2\pi},1)+o(\Delta_{\rm total}^n)
\ee
In particular this means that for $\Delta_{\rm total}$ large enough,
$D$ will only give positive contributions, no matter which
massive representation we consider. Since the total contribution
of $Z_{\ml\bar{\ma}}$ must vanish, it follows that there must
be at least some massive fields at small $\Delta$.

More precisely, let $\Delta_0^{Q\bar Q}$ be the highest
root of the polynomial $P_D(\ma^Q_\Delta\bar\ma^{\bar Q}_{\bar{\Delta}})$.
Defining
\be
\Delta_0=\max_{Q,\bar Q}(\Delta_0^{Q\bar Q})\ ,
\ee
it follows that there must be at least one massive
representation with
\be
\Delta_{\rm total} < \Delta_0\ .
\ee
This is the central idea of the method. Similar statements 
can of course be made if we consider holomorphic
operators $\partial_s$ instead, in which case we can
obtain bounds on $\Delta$ instead of $\Delta_{\rm total}$.

To get the strongest bound, we should find an operator
whose highest roots $\Delta_0^{ab}$ are as low as possible.
Generically, the more roots there are, the bigger the highest
roots will be. This suggests that we should first look
at operators of low degree, \eg with only one derivative
in $s,\bs$. It is thus natural to concentrate on the operator
\be \label{dsdw2}
D= -\partial_{s+\bs} + A(h^{Q,\bar Q},N^\Delta_{Q\bar Q})\partial_w^2\ ,
\ee
where $A$ is a function of the Hodge numbers $h^{Q,\bar Q}$
and the coefficients $N^\Delta_{Q\bar Q}$ such that $D$ annihilates
the massless part of the partition function. By construction it is clear that $A$
is the quotient of two linear functions in $h$ and $N$.
Acting on the massive representations gives linear
polynomials in $\Delta_{\rm total}$
\be
P_D(\ma^Q_\Delta\bar\ma^{\bar Q}_{\bar{\Delta}})=u^{Q\bar Q}\Delta_{\rm total}+v^{Q\bar Q} + A(h,N)w^{Q\bar Q}\ ,
\ee 
whose roots are given by
\be\label{dsdw2root}
\Delta^{Q\bar Q}_0 = -A(h,N)\frac{w^{Q\bar Q}}{u^{Q\bar Q}} -\frac{v^{Q\bar Q}}{u^{Q\bar Q}}\ .
\ee
From the arguments above we know that $u^{Q\bar Q}>0$. The sign
of $v^{Q\bar Q}$ can vary, depending on how many states of weight
smaller than $\Delta+c/24$ there are. Similarly the sign of
$w^{Q\bar Q}$ depends on the number of charged states in
the representation\footnote{Although it may seem that $\partial_w^2$
always gives negative contributions, it is important to remember the
prefactor introduced in (\ref{hatdef}), which gives a positive
contribution.}.

For fixed Hodge number and $N^\Delta_{Q\bar Q}$, we thus obtain
an upper bound $\Delta_0$ on the lowest lying massive field.
As we discussed in the introduction, we are often interested 
in constraints for Calabi-Yau with large
Hodge numbers. Since $A$ is the ratio of two linear functions,
it will tend to some finite asymptotic value. This means that
even in the limit of very large Hodge numbers we can still get
meaningful bounds.


\subsection{Calabi-Yau threefolds}
\begin{figure}[htbp]
\begin{center}
\includegraphics{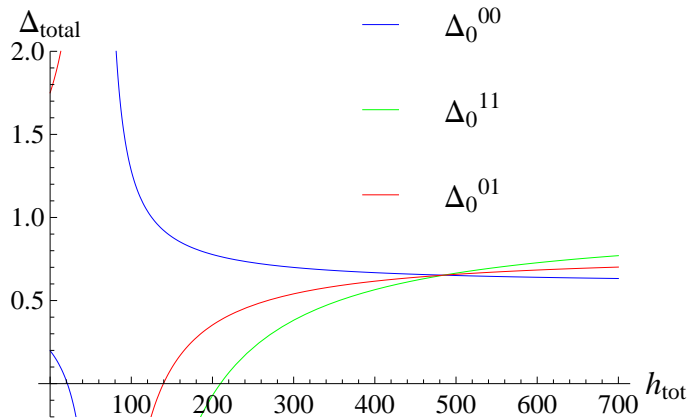}
\caption{$\Delta_0^{Q\bar Q}$ for the three combinations of massive representations. 
}
\label{Fig:c9constraints}
\end{center}
\end{figure}

Let us discuss this for the case of Calabi-Yau 3-folds. 
First we want to construct an operator $D$ as in
(\ref{dsdw2}).
Defining 
\begin{eqnarray*}
\partial_{s+\bs}\hat \ml^0 \bar{\hat \ml}^0 \bigr|_0= \alpha_0 \ ,&&
\partial_{w}^2\hat \ml^0 \bar{\hat \ml}^0 \bigr|_0= \beta_0\ ,\\
\partial_{s+\bs}\hat \ml^1 \bar{\hat \ml}^1 \bigr|_0= -\alpha_1 \ ,&&
\partial_{w}^2\hat \ml^1 \bar{\hat \ml}^1 \bigr|_0= -\beta_1\ ,
\end{eqnarray*}
the differential operator
\be\label{Dsw2}
D=-\partial_{s+\bs}+\frac{\alpha_0-2\alpha_1 h_{\rm total}}{\beta_0-2\beta_1 h_{\rm total}}
\partial_w^2
\ee
with $h_{\rm total}=h^{1,1}+h^{2,1}$ annihilates 
$Z_{\ml\bar{\ml}} +Z_{\ma\bar{\ml}}+Z_{\ml\bar{\ma}}$. The
numerical values of the coefficients are
\be
\alpha_0= 523.5\ldots,\qquad \alpha_1 = 0.53\ldots,
\qquad\beta_0=1046.5\ldots,\qquad \beta_1= 8.24\ldots .
\ee
We now need to investigate the action of the differential
operator (\ref{Dsw2}) on the massive representations
$\ma^0_\Delta\bar{\ma}^0_{\bar\Delta}$, $\ma^1_\Delta\bar{\ma}^1_{\bar\Delta}$
and $\ma^0_\Delta\bar{\ma}^1_{\bar\Delta}+\ma^1_\Delta\bar{\ma}^0_{\bar\Delta}$.
Applying (\ref{dsdw2root}) gives
\be
\Delta^{Q\bar Q}_0 = -\frac{\alpha_0-2\alpha_1 h_{\rm total}}{\beta_0-2\beta_1 h_{\rm total}}\frac{w^{Q\bar Q}}{u^{Q\bar Q}} -\frac{v^{Q\bar Q}}{u^{Q\bar Q}}\ .
\ee
The $\Delta_0^{Q\bar Q}$ are plotted in figure~\ref{Fig:c9constraints}.

Let us briefly discuss this result.
As a function of the Hodge numbers $h_{\rm total}$, the root $\Delta^{ab}_0$
is a hyperboloid with singularity between $h_{\rm total}=63$ and 64, which
means that the bound is worst around these values of the
Hodge number.
For large Hodge numbers it asymptotes to
\be\label{CY3asym}
\Delta^{Q\bar Q}_0 \rightarrow -\frac{\frac{\alpha_1}{\beta_1}w^{Q\bar Q}+v^{Q\bar Q}}{u^{Q\bar Q}}\ .
\ee
Since the lowest lying states $\ma^0$ have charge 0,
it is not surprising that $w^{00}>0$, so that
$\Delta_0^{00}$ approaches its asymptotic value
from above. The other representations have low lying
states of charge 1, so that $w^{11},w^{01}<0$.
Note that from (\ref{massivetomassless}) it follows that
\be
w^{11}= -4e^{2\pi}\beta_1\ , \qquad u^{11}+v^{11}=4e^{2\pi}\alpha_1\ ,
\ee
so that $\Delta_0^{11} \rightarrow 1$.
This tells us that 
\be
\Delta_0^{11}<1 \qquad \textrm{for}\qquad h_{\rm total}\geq 64\ .
\ee
On the other hand unitarity imposes that for $\ma^1_\Delta\bar\ma^1_{\bar{\Delta}}$ 
the weights have to satisfy
$\Delta,\bar{\Delta} \geq \frac{1}{2}$. It follows that this
type of massive representation always gives positive 
contributions for large Hodge numbers.
A similar argument shows that $\Delta_0^{01}$
actually cannot contribute negatively for $64 \leq h_{\rm total} \leq 267$.

To summarize, the differential operator (\ref{dsdw2}) allowed
us to find bounds on the lowest lying massive representation.
In particular this bound is also applicable for Hodge numbers
smaller than 492. Unfortunately it becomes weaker and weaker
as we approach $h_{\rm total}=64$. To remedy this, let us now
turn to another type of differential operator.

\subsection{Constraints from the ordinary partition function}
Let us discuss what happens if we apply the methods
of \cite{Hellerman:2009bu} directly to our setup. Instead of 
(\ref{dsdw2}) we will instead use the operator
\be\label{dsds3}
D= \partial_{s+\bs}^3 - A(h^{Q,\bar Q},N^\Delta_{Q\bar Q})\partial_{s+\bs}\ .
\ee
Again we fix $A$ in such a way that the massless part of the
partition function is annihilated.
Since (\ref{dsds3}) is a third order operator, $P_D$ has
three roots, the bound $\Delta_0^{Q\bar Q}$ being the largest of the three.
It turns out that for all three combinations of massive representations
those three roots are real for $h_{\rm total}\leq 492$.
For $h_{\rm total}\geq 493$, two of the roots become imaginary, so that
the bound jumps to a much lower value. For very small 
Hodge numbers however this method gives the most
stringent bound, as shown in figure 4. 

\begin{figure}[htbp]
\begin{center}
\includegraphics{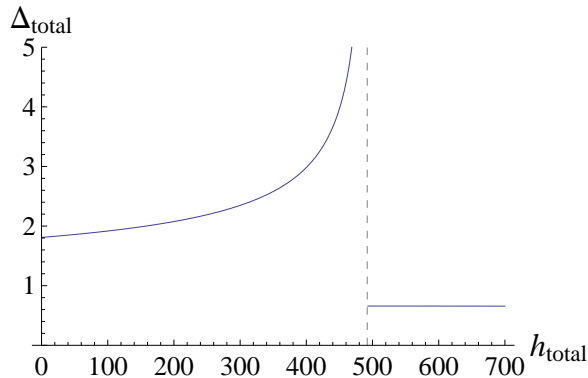}
\caption{$\Delta_0^{Q\bar Q}$ for the various
combinations of massive representations.  }
\label{Fig:c9constraintsds3}
\end{center}
\end{figure}

Collecting our results thus gives the following picture:
for $h_{\rm total}\leq 82$, the differential operator
(\ref{dsds3}) gives the best bound. In the range
$83 \leq h_{\rm total} \leq 492$, the best bound comes from
(\ref{Dsw2}). For $h_{\rm total}$ larger than that, 
the best bound is obtained from (\ref{Nbound}).
This is summarized in figure 1 in the introduction section of this paper.

\subsection{Enhanced symmetries}
So far we have always assumed that $Z_{cross}$ contains
no vacuum terms. If there is a term of the form
$\ml^0\bar{\ma}^{\bar Q}_{\bar \Delta}$, $\bar \Delta \in \frac{1}{2}\Z$ then the conformal field theory
has an enhanced symmetry, since there is then a field of
weight $(0,\bar\Delta)$ in the theory. 
Such symmetries can come from instance 
of isometries of the underlying Calabi-Yau, even
though this is only possible in the non-compact case. Especially
for higher spin symmetries there is in general no simple 
geometric interpretation. At a generic point in moduli space
one does not expect enhanced symmetry, but there may 
be points of symmetry enhancement. 

\begin{figure}[htbp]
\begin{center}
\includegraphics{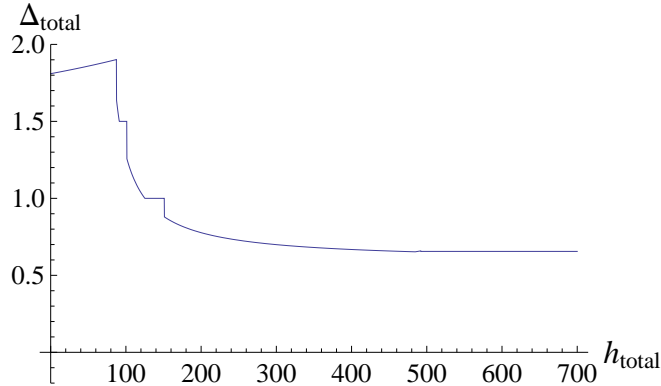}
\caption{Upper bound for the total weight $\Delta_{\rm total}$ of the 
lowest lying massive field for a Calabi-Yau threefold
with enhanced symmetry.}
\label{Fig:c9constraintsEnhanced}
\end{center}
\end{figure}

It is straightforward to repeat our analysis in such a case
by generalizing the ansatz for $Z_{\ml\bar\ma}$. 
The result does not differ by much, as figure~\ref{Fig:c9constraintsEnhanced} shows.
The bound for large Hodge numbers remains unaffected,
and only for very specific Hodge numbers can 
fields of weight $(0,1)$ and $(0,\frac{3}{2})$ weaken
the bound slightly.

In such cases however one should clearly analyze the enhanced
symmetry in more detail, and compute the characters
with respect to the bigger symmetry of the theory.
Repeating our type of analysis for these representations
should lead to even stronger constraints.


\subsection{Other dimensions}
In principle we can apply the same type of analysis 
to higher dimensional Calabi-Yau manifolds.
The situation is completely analog to what we have discussed,
but the analysis will be more involved since there are more
massive representations. Moreover the differential operator
(\ref{dsdw2}) will now depend on various linear combinations
of the Hodge numbers as discussed in section~\ref{ss:dfold}
and in general also on the $N^\Delta_{Q\bar Q}$.

On the other hand we can certainly repeat the analysis
for K3. In this case $Z_0$ is completely fixed, since
the Hodge numbers of K3 are determined.
In section~\ref{ss:CYpf} we described how to obtain 
$Z_{\ml\bar\ma}$ from the elliptic genus.
It turns out however that here $P_{\partial_{w}^2}$ vanishes
for the massive representation, and therefore also for the
the total contribution of the massless representations.
This means that we cannot construct an operator of the
form (\ref{Dsw2}). Although it is possible to construct
operators with higher derivatives in $w$, it turns out that
the best bound is obtained by an operator of the form
(\ref{dsds3}), which gives 
\be
\Delta_0 = 1.147\ldots\ .
\ee

\medskip
\noindent {\bf Acknowledgment}

We thank T.~Eguchi, M.~Freedman, M.~Gaberdiel, S.~Hellerman, C.~Schmidt-Colinet, W.~Taylor, C.~Vafa and S.-T.~Yau for useful
discussions. We thank D.~Friedan and T.~H\"ubsch for their comments on the earlier version of the paper. We thank the hospitality of 
the Aspen Center for Physics (NSF grant 1066293) and the Simons Center for
Geometry and Physics. This work 
is supported in part by U.S. DOE grant DE-FG03-92-ER40701.
The work of CAK is also supported in part by the John A. McCone Postdoctoral Fellowship at Caltech
and U.S. DOE grant DE-FG02-96ER40959. 
The work of HO is also  
supported in part by a Simons Investigator award from the Simons Foundation,
the Fred Kavli Professorship at Caltech,  
and by the WPI Initiative of MEXT of Japan
and JSPS Grant-in-Aid for Scientific Research C-23540285.

\appendix

\section{Representations of the extended $N=2$ superconformal algebra}\label{app:N2char}
A worldsheet theory for Calabi-Yau compactification of
type II superstring theory without flux or for $N=(2,2)$ compactification
of heterotic string theory 
has $N=2$ superconformal symmetry.
In addition we also want symmetry under spectral flow.
More precisely, spectral flow by one unit will map
the Ramond (R) and Neveu-Schwarz (NS) sectors to themselves, and the corresponding
operators will be part of the symmetry algebra.
Another way to say this is that a CY $d$-fold has
a holomorphic $(d,0)$ form $\Omega$. The most efficient
way to deal with the representations is to write
them as representations of this bigger symmetry algebra.
For K3 the forms are the currents $J^\pm$ which enhance
the symmetries to $N=4$. 

For a Calabi-Yau $d$-fold the extended $N=2$ superconformal
algebra has central charge $c=3d$. Let us define
$k:=d-1$. 
\begin{description}
\item[Massive representations:] There are $d-1$ massive representations
\be
\ma^Q_h\qquad Q=0,\pm1,\ldots, \pm\left(\lfloor d/2\rfloor-1\right),\frac{d-1}{2}\ ,
\ee
where the last representation only appears for $d$ odd.
They are given by
\be
\ma^Q_\Delta(q,y) = q^\Delta F_{NS}(q,y)q^{-k/8-Q^2/2k-1/8}\sum_{m\in\Z}q^{\frac{k}{2}(m+Q/k)^2}y^{km+Q}\ .
\ee
Note that 
\be
\ma^{-Q}_\Delta(q,y) = \ma^Q_\Delta(q,y^{-1})\ .
\ee
\item[Massless representations:]
There are $d$ massless representations,
\be
\ml^Q\qquad Q=0,\pm1,\ldots \lfloor \frac{d-1}{2}\rfloor, \frac{d}{2}\ ,
\ee
where the last representation only appears for $d$ even.
For $Q>0$ they are given by
\be
\ml^Q(q,y) = q^{-1/8}F_{NS}(q,y)\sum_{m\in\Z}\frac{(yq^{m+1/2})^{Q-k/2}}{1+yq^{m+1/2}}q^{\frac{k}{2}(m+1/2)^2}y^{k(m+1/2)}\ .
\ee
For $Q<0$ we define
\be
\ml^Q(q,y) = \ml^{-Q}(q,y^{-1})\ .
\ee
The vacuum representation with $Q=0$ is given by
\be
\ml^0(q,y) = q^{-5/8-1/2k - k/8}F_{NS}(q,y)\sum_{m\in\Z}
\frac{(1-q)q^{\frac{k}{2}(m + 1/k)^2}y^{km+1}}{(1+yq^{m-1/2})(1+yq^{m+1/2})}\ .
\ee
\end{description}
Here $F_{NS}$ denotes the partition function of $N=2$ superconformal
descendants without any additional relations,
\be
F_{NS}(y,q)=\prod_{n\geq1}\frac{(1+yq^{n-1/2})(1+y^{-1}q^{n-1/2})}{(1-q^n)^2}\ .
\ee
Note that from our definitions
\be
\ml^{-Q}=\ml^{d-Q}\qquad \textrm{for} \ 0<Q<d/2\ .
\ee
It will thus often be useful to label
the massless representations by
\be
\ml^Q\qquad Q=0,1\ldots, d-1
\ee
instead.
The Witten index of the representations is given by
\be
ind = \left\{\begin{array}{cl}0&\textrm{massive}\\
(-1)^{d-Q}& \textrm{massless}\ Q=1,\ldots, d-1\\
1+(-1)^d&\textrm{massless}\ Q = 0 \end{array}\right.
\ee
When the massive representations reach the unitarity
bound $\Delta=\frac{1}{2}|Q|$, they become reducible and decompose
into massive representations as
\begin{eqnarray}
\ma^Q_{Q/2}= \ml^Q+\ml^{Q+1} && Q>0\nonumber\\
\ma^{-Q}_{Q/2}= \ml^{-Q}+\ml^{-Q-1} && Q>0\\
\ma^0_0= \ml^0 +\ml^1+\ml^{-1} && Q=0\nonumber\ .
\end{eqnarray}

\section{Elliptic genus}\label{app:ellGenus}
In general the elliptic genus of a CY $d$-fold is given by
a weak Jacobi form of weight 0 and index $d/2$.
For the definition and a general introduction to weak Jacobi forms see \cite{EichlerZagier}.
What is important for us is that
the ring of weak Jacobi forms $J_{2k,m}$ of integer index $m$
and even weight $2k$ is generated by $E_4,E_6,\phi_{0,1},\phi_{-2,1}$.
For explicit expressions and a discussion of this basis see for
example \cite{Gaberdiel:2008xb}, and for its application
to the topology and geometry of Calabi-Yau manifolds see \cite{Gritsenko:1999fk}. 
For half-integer index on the other hand we have
\be
J_{2k,m+1/2}=\phi_{0,3/2}\cdot J_{2k,m-1}\ ,\qquad J_{2k+1,m+1/2}=\phi_{-1,1/2}\cdot J_{2k+2,m}
\ee
where
\begin{eqnarray}
\phi_{0,3/2}(\tau,z)&=&y^{-1/2}\prod_{n\geq1}(1 + q^{n-1}y)(1 + q^{n}y^{-1})
(1 - q^{2n-1}y^2)(1 - q^{2n-1}y^{-2})\\
\phi_{-1,1/2}(\tau,z)&=&-y^{-1/2}\prod_{n\geq1}(1 - q^{n-1}y)(1 - q^ny^{-1})(1 - qn)^{-2} \ .
\end{eqnarray}
In particular it follows that the elliptic genus of a Calabi-Yau 3-fold is 
given a multiple of $\phi_{0,3/2}$, which gives 1 constraint on
the Hodge numbers \cite{Gritsenko:1999fk}.

We can relate the coefficients
of the elliptic genus $J$ of a Calabi Yau manifold of 
complex dimension $d$ to its Hodge numbers by
\be
J(q,y)= \sum_{p=0}^d (-1)^p \chi_p(M) y^{\frac{d}{2}-p} + O(q)
\ee
with $\chi_p(M)= \sum_q (-1)^q h^{p,q}(M)$.

\end{document}